\documentclass[prl,aps,twocolumn,showpacs,superscriptaddress]{revtex4}
\usepackage{epsfig,latexsym, graphicx}

\def\Re{\rm Re}
\def\Im{\rm Im}
\begin{document}
\title{Clauser-Horne-Bell inequality for three three-dimensional systems}

\author{Jing-Ling Chen}
\affiliation{Laboratory of Computational Physics, Institute of
Applied Physics and Computational Mathematics,  P.O. Box 8009(26),
Beijing 100088, People's Republic of China}
\author{Dagomir Kaszlikowski}
\affiliation{Department of Physics, National University of
Singapore, 10 Kent Ridge Crescent, Singapore 119260}
\affiliation{Instytut Fizyki Teoretycznej i Astrofizyki,
Uniwersytet Gda\'nski, PL-80-952, Gda\'nsk, Poland}
\author{L. C. Kwek}
\affiliation{Department of Physics, National University of
Singapore, 10 Kent Ridge Crescent, Singapore 119260}
\affiliation{National Institute of Education, Nanyang
Technological University, 1 Nanyang Walk, Singapore 639798}
\author{C. H. Oh}
\affiliation{Department of Physics, National University of
Singapore, 10 Kent Ridge Crescent, Singapore 119260}

\begin{abstract}
In this paper we show a Bell inequality of Clauser-Horne type
for three three-dimensional systems (qutrits). Violation of the
inequality by quantum mechanics is shown for the case in which
each of the three observers measures two non-commuting
observables, defined by the so called unbiased symmetric six port
beamsplitters, on a maximally entangled state of three qutrits. The strength
of the violation of the inequality agrees with 
the numerical results presented in {\it Kaszlikowski et. al.}, quant-ph//0202019.
\end{abstract}

\maketitle
It has recently been found that two entangled $N$-dimensional
systems (qu$N$its) generate correlations that are more
robust against noise than those generated by two entangled qubits
\cite{KASZPRL,DURT,CHQ,POP}. Moreover, this robustness increases
with the dimension $N$. It has also been shown numerically
\cite{KGHZ} that a similar trend holds in the case of three
entangled quantum systems, in the sense that correlations generated by
three qutrits are more robust against noise than in the case of three qubits. 

There is a substantial difference between the correlations
generated by two and three entangled systems. In the case of
bipartite entanglement the strentgh of violation of Bell inequalities 
(defined as the minimal amount of noise that is necessary to hide 
non-classsical nature of quantum correlations)
for the maximally entangled state increases with the dimension
of entangled objects. Whereas in the tripartite qutrit entanglement, as
it has been shown in Ref. \cite{KGHZ}, maximally entangled state is
less robust against noise than its qubit analogon (tripartite
maximally entangled state of qubits is state that is the most robust 
against noise). 
However, there exists a non-maximally entangled 
state of three qutrits, which is much more robust against noise
than maximally entangled state of three qubits (for more detailed discussion
see Ref. \cite{KGHZ}).

To understand this peculiar behaviour (peculiar because one would expect that
it should be maximally entangled state that is more robust against noise rather than
non-maximally entangled one; similar phenomena has been shown in \cite{gisin} for
two entangled qutrits) of tripartite qutrit entanglement
one needs Bell inequalities suitable for this problem. This inequalities
must be able at least to reproduce the numerical results 
in Ref. \cite{KGHZ}.   
Note, that such inequalities can be also useful in analysis of other interesting
problems like, for instance, tripartite
bound entanglement. They also provide us with entanglement witness which is one of  
tools in research conerning entanglement.

In this paper, we present a Clauser-Horne-Bell inequality for
three qutrits for the situation in which each observer measures
two non-commuting observables. This inequality imposes necessary
conditions on the existence of local realistic description of the
correlations generated by three qutrits.

We show the violation of this inequality by quantum mechanics in a
gedanken experiment in which observables measured by the observers
are defined by unbiased symmetric six-port beamsplitters on
maximally entangled state. The violation is the same as predicted
numerically in Ref. \cite{KGHZ}. 

Let us consider a Bell-type gedanken experiment with three
observers each measuring two observables on some quantum state
$\hat{\rho}$. We denote the observables by $\hat{A}_1, \hat{A}_2$ for
the first observer (Alice), $\hat{B}_1,\hat{B}_2$ for the second
observer (Bob) and $\hat{C}_1,\hat{C}_2$ for the third one
(Celine). The measurement of each observable yields three distinct
outcomes (numbers) which we denote by $a_1^{i},a_2^{i},a_3^{i}$ for Alice's
measurement of the observable $\hat{A}_i$, $b_1^{j},b_2^j,b_3^j$
for Bob's measurement of the observable $\hat{B}_j$ and
$c_1^k,c_2^k,c_3^k$ for Celine's measurement of the observable
$\hat{C}_k$ ($i,j,k=1,2$). Specifically, the observable
$\hat{A}_i$ has the spectral decomposition $\hat{A}_i =
a_1^{i}\hat{P}^i_1+a_2^{i}\hat{P}^i_2+a_3^{i}\hat{P}^i_3$, where
$\hat{P}^i_1,\hat{P}^i_2,\hat{P}^i_3$ are mutually orthogonal
projectors.  Similarly, the observable $\hat{B}_j$ has the
spectral decomposition $\hat{B}_j =
b_1^{j}\hat{Q}^j_1+b_2^{j}\hat{Q}^j_2+b_3^{j}\hat{Q}^j_3$ and the
observable $\hat{C}_k =
c_1^{k}\hat{R}^k_1+c_2^{k}\hat{R}^k_2+c_3^{k}\hat{R}^k_3$ where
$\hat{Q}^j_\zeta$ as well as $\hat{R}^k_\zeta$ ($\zeta= 1, 2, 3$) are
mutually orthogonal projectors.

The probability of obtaining the set of three numbers
$(a_{l_i}^i,b_{m_j}^j,c_{n_k}^k)$ in a simultaneous measurement of
observables $\hat{A}_i,\hat{B}_j,\hat{C}_k$ on the state $\rho$ is
denoted by $W_{QM}(a_{l_i}^i,b_{m_j}^j,c_{n_k}^k)$, where $l_i,
m_j, n_k$ assumes the values $1, 2, 3$, and is given by the standard
formula
\begin{equation}
W_{QM}(a_{l_i}^i,b_{m_j}^j,c_{n_k}^k) = Tr(\rho\hat{P}^i_{l_i}
\otimes\hat{Q}^j_{m_j}\otimes\hat{R}^k_{n_k}). \label{prob}
\end{equation}
According to quantum theory, everything that can be measured in
this gedanken experiment is given by the set of these $8\times 27
= 216$ probabilities.

Local realistic (classical) description of the above experiment is
equivalent to the existence of a joint probability distribution
from which the quantum probabilities
$W_{QM}(a_{l_i}^i,b_{m_j}^j,c_{n_k}^k)$ can be derived as the
marginals. Let us denote this hypothetical joint probability
distribution by
$W_{LR}(a^1_{l_1},a^2_{l_2};b^1_{m_1},b^2_{m_2};c^1_{n_1},c^2_{n_2})$.
Thus, a local realistic description of the experiment exists {\it if
and only if} the following marginals

\begin{widetext}
\begin{eqnarray}
&&W_{LR}(a_{l_i}^i,b_{m_j}^j,c_{n_k}^k)
=\sum_{l_{i+1}=1}^{3}\sum_{m_{j+1}=1}^3\sum_{n_{k+1}=1}^{3}
W_{LR}(a^1_{l_1},a^2_{l_2};b^1_{m_1},b^2_{m_2};c^1_{n_1},c^2_{n_2})
\label{marg}
\end{eqnarray}
\end{widetext}
are equal to the quantum probabilities, i.e.,
$W_{LR}(a_{l_i}^i,b_{m_j}^j,c_{n_k}^k)=W_{QM}(a_{l_i}^i,b_{m_j}^j,c_{n_k}^k)$
where the addition on the indices is computed using modulo 2
arithmetics.

Owing to (\ref{marg}), $W_{LR}(a_l^i,b_m^j,c_n^k)$ must obey the
following inequality
\begin{widetext}
\begin{eqnarray}
&&-\Gamma_{221}-\Gamma_{111}+2\Gamma_{122}+\Gamma^{'}_{121}-\Gamma^{'}_{212}+
\Gamma^{''}_{211}+\Gamma^{''}_{222}+\Gamma^{''}_{112}\leq 3,
\end{eqnarray}
where
\end{widetext}
\begin{eqnarray}
&&\Gamma_{ijk} = \sum_{l+m+n=1~mod~3}W_{LR}(a_l^i,b_m^j,c_n^k)\nonumber\\
&&\Gamma^{'}_{i'j'k'} = \sum_{l+m+n= 0~mod~3}W_{LR}(a_1^{i'},b_1^{j'},c_2^{k'})\nonumber\\
&&\Gamma^{''}_{i''j''k''} = \sum_{l+m+n = 2~mod~3}W_{LR}(a_1^{i''},b_1^{j''},c_2^{k''}), 
\label{ineq}
\end{eqnarray}
and $(i,j,k)=(221,111,122)$, $(i',j',k')=(121,212)$, $(i'',j'',k'')=(211,222,112)$.
This is the Clauser-Horne-Bell inequality for three qutrits. It
must be obeyed by any local realistic theory that claims to
reproduce the correlations generated by three qutrits.

To prove the inequality in (\ref{ineq}), we first replace the
marginals in the left hand side of the inequality in (\ref{ineq})
by the appropriate sums of joint probabilities given in
(\ref{marg}). Naturally, we get an extremely long expression (this is why
we do not show it here) in
which the joint probabilities $W_{LR}(a_{l_1}^{1}, a_{l_2}^{2};
b_{m_1}^{1}, b_{m_2}^{2}; c_{n_1}^{1}, c_{n_2}^{2})$ appear only
with coefficients -3, 0 or 3 and nothing else.  Since the sum of
all joint probabilities adds to one, i.e., $\sum_{a_{l_1}^{1}, a_{l_2}^{2},
b_{m_1}^{1}, b_{m_2}^{2}, c_{n_1}^{1}, c_{n_2}^{2}=1}^{3} W_{LR}(a_{l_1}^{1}, a_{l_2}^{2};
b_{m_1}^{1}, b_{m_2}^{2}; c_{n_1}^{1}, c_{n_2}^{2})\linebreak= 1$, 
it follows immediately that the entire expression is less than or equal to 3.

We should stress at this point that the above inequality is a member
of the set of inequalities that can obtained from (\ref{ineq}) by permutations of
indices enumerating the outcomes of the measurements as well as the permutations of
indices enumerating the observables.

Consider a gedanken experiment in which Alice, Bob and Celine
measure observables defined by unbiased symmetric six-port
beamsplitters \cite{tritters} on the maximally entangled state of
three qutrits $|\psi\rangle = {1\over\sqrt 3}(|111\rangle
+|222\rangle + |333\rangle)$.

The unbiased symmetric six-port beamsplitters is an optical device
with three input and three output ports. In front of every input
port there is a phase shifter that changes the phase of the photon
entering the given port. If a phase shifter in some input port is
set to zero and a photon enters the device through this port then
it has an equal chance of leaving the device through any output
port. The phase shifters can be changed by the observers; they
represent the local macroscopic parameters available to the
observers.

The matrix elements of an unbiased symmetric six-port beamsplitter
are given by $U_{kl}(\vec{\phi})={1\over\sqrt
3}\alpha^{(k-1)(l-1)}\exp(i\phi_l)$, where $\vec{\phi}
=(\phi_1,\phi_2,\phi_3)$ and $\phi_l$ ($k,l=1,2,3$) are the settings
of the appropriate phase shifters (for convenience we denote them
as a three dimensional vector $\vec{\phi}$) and
$\alpha=\exp({2i\pi\over 3})$.

The observables measured by Alice, Bob and Celine are now defined
as follows. The set of projectors for Alice's $i$-th experiment is
given by $\hat{P}^i_l=U_A^{\dagger}(\vec{\phi}_i)|l\rangle\langle
l|U_A(\vec{\phi}_i)$ ($l=1,2,3$), where $U_A(\vec{\phi}_i)$ is the
matrix of Alice's unbiased symmetric six-port beamsplitter defined
by the set of phases $\vec{\phi}_i=(\phi^i_1,\phi^i_2,\phi^i_3)$,
Bob's set of projectors $j$-th experiment is given by
$\hat{Q}^j_m=U_B^{\dagger}(\vec{\psi}_j)|m\rangle\langle
m|U_B(\vec{\psi}_j)$, where
$\vec{\psi}_j=(\psi^j_1,\psi^j_2,\psi^j_3)$ is a set of Bob's
phases defining his unbiased symmetric six-port beamsplitter,
whereas Celine's projectors in the $k$-th experiment is given by
$\hat{R}^k_n=U_C^{\dagger}(\vec{\delta}_k)|n\rangle\langle
n|U_C(\vec{\delta}_k)$, where
$\vec{\delta}_k=(\delta^k_1,\delta^k_2,\delta_3^k)$ is a set of
Celine's phases defining her unbiased symmetric six-port
beamsplitter.

To each result of the measurement of the projectors
$\hat{P}^i_n,\hat{Q}^j_n,\hat{R}^k_n$ for any $i,j,k$ we ascribe
the complex number $\alpha^n$ ($n=1,2,3$), namely $a_{l_1}^1,
a_{l_2}^2, b_{m_1}^1, \dots $  have been assigned the values
$\alpha^{l_1}, \alpha^{l_2}, \alpha^{m_1}, \dots$ respectively.
This special assignment was first used in Ref. \cite{tritters} to
generalize the Bell experiment for higher dimensions.

In this way, the probability of getting the set of three numbers
$(a_{l_i}^i,b_{m_j}^j,c_{n_k}^k)$ can now be computed using the
formula (\ref{prob}). Because we are not going to use the explicit
form of these probabilities, interested readers are kindly
referred to Ref. \cite{tritters,KGHZ}. However, note that we need
to use the following property regarding these probabilities. All
the probabilities $W_{QM}(a_{l_i}^i,b_{m_j}^j,c_{n_k}^k)$ can be
sorted into three groups consisting of nine equal probabilities.
The first group consists of the probabilities for which
$l_i+m_j+n_k=1$ mod $3$, the second one consists of the
probabilities for which $l_i+m_j+n_k=2$ mod $3$ and the third one
consists of the probabilities for which $l_i+m_j+n_k=0$ mod $3$.
Let us denote each probability (they are equal, so it suffices to
take an arbitrary one as a representative of the whole group) from
the first group by $W_{QM}^1(ijk)$, from the second one by
$W_{QM}^2(ijk)$ and from the third one by $W_{QM}^3(ijk)$. It is
obvious that we have $W_{QM}^1(ijk)+W_{QM}^2(ijk)+W_{QM}^3(ijk) =
{1\over 9}$ for any triple $i,j,k$.

Let us now define the following correlation function (for details
see \cite{tritters}) for each triple of experiments that we denote
by $Q_{ijk}$
\begin{eqnarray}
&&Q_{ijk} = \sum_{l_i,m_j,n_k=1}^3
\alpha^{l_i+m_j+n_k}W_{QM}(a_{l_i}^i,b_{m_j}^j,c_{n_k}^k).
\end{eqnarray}
Using the explicit form of the probabilities, it can be shown
easily that such correlation function acquires the following
symmetric form
\begin{widetext}
\begin{eqnarray}
&&Q_{ijk} = {1\over 3}(\exp(\phi^i_1-\phi^i_2+\psi^j_1-\phi^j_2+\delta^k_1-\delta^k_2)+
\exp(\phi^i_2-\phi^i_3+\psi^j_2-\phi^j_3+\delta^k_2-\delta^k_3)\nonumber\\
&&+\exp(\phi^i_3-\phi^i_1+\psi^j_3-\phi^j_1+\delta^k_3-\delta^k_1)).
\end{eqnarray}
\end{widetext}

The splitting of the probabilities into the three groups implies
that this correlation function conveys as much information about
the experiment as the probabilities themselves. In fact, there is
a one-to-one mapping between the correlation function and the
probabilities so that the following equations hold
\begin{eqnarray}
&&W_{QM}^1(ijk) = {1\over 27} (1-\Re Q_{ijk}+\sqrt 3 \Im Q_{ijk})\nonumber\\
&&W_{QM}^2(ijk) = {1\over 27} (1-\Re Q_{ijk}-\sqrt 3 \Im Q_{ijk})\nonumber\\
&&W_{QM}^3(ijk) = {1\over 9} -W_{QM}^1(ijk)-W_{QM}^2(ijk).
\label{trick}
\end{eqnarray}

\noindent Putting the probabilities expressed by the equations
(\ref{trick}) into the Clauser-Horne-Bell inequality (\ref{ineq}),
we obtain the following inequality (which is totally equivalent to
(\ref{ineq}) in the case considered here)
\begin{widetext}
\begin{equation}
\Re[Q_{121} - Q_{212} + \alpha \left( Q_{112} + Q_{211} + Q_{222}
\right)+ \alpha^2 \left( 2 Q_{122} - Q_{111} - Q_{221} \right)]
\leq 3 \label{chsh}
\end{equation}
\end{widetext}

We will show that this inequality is not satisfied by quantum
mechanics for appropriate choice of the phase shifts for Alice,
Bob and Celine. The phase shifts are as follows $ \vec{\phi}_1 =
(0,0,{2\pi\over 3}), \vec{\phi}_2 = (0,0,0), \vec{\psi}_1 =
(0,0,\pi), \vec{\psi}_2 = (0,0,{5\pi\over 3}), \vec{\delta}_1 =
(0,{\pi\over 3},0), \vec{\delta}_2 = (0,\pi,0).$ The values of the
correlation function computed using the above phase shifts read
$Q_{111}={1\over 3}(1+\alpha^2), Q_{112}={2\over
3}\alpha^2,Q_{121} = {2\over 3}, Q_{122} = -{2\over
3}(1+\alpha^2), Q_{211} = {2\over 3}\alpha^2, Q_{212} = -{1\over
3}, Q_{221} = -{1\over 3}\alpha, Q_{222} = {2\over 3}\alpha^2$.
Putting them into the left hand side of the inequality in
(\ref{chsh}) we arrive at a violation of the inequality in which
the left hand side is equal to 5.

In Ref. \cite{KASZPRL}, a proposal was made to measure the
strength of violation of local realism by the minimal amount of
noise that must be added to the system in order to hide the
non-classical character of the observed correlations. This is
equivalent to a replacement of the pure state
$|\psi\rangle\langle\psi|$ by the mixed state $\rho(F)$ of the
form $\rho(F)= (1-F)|\psi\rangle\langle\psi| + {F\over 27}I\otimes
I\otimes I$, where $I$ is an identity matrix and where $F$ ($0\leq
F\leq 1$) is the amount of noise present in the system.

It can be checked immediately that such addition of the noise in
the gedenken experiment considered here changes the correlation
function $Q_{ijk}$ to $Q_{ijk}^F = (1-F)Q_{ijk}$. Therefore, the
minimal amount of noise $F_{min}$ that must be added to the system to
conceal the non-classicality of quantum correlations is $F_{min} =
{4\over 10}$, which is consistent with the numerical results
presented in Ref. \cite{KGHZ}.


In conclusion, we have found the Clauser-Horne-Bell type
inequality (we name it in this way because the inequality
involves probabilities and not a correlation function of some sort) 
for three qutrits that gives us the necessary
conditions for the existence of a local realistic (classical)
description of quantum correlations. We have shown that the
violation of this inequality for the gedenken experiment with the
maximally entangled state in which two observables defined by
unbiased symmetric six-port beamsplitters are measured at three
sites. The strength of the violation agrees with the numerical
results obtained in Ref. \cite{KGHZ}. Moreover, the numerical
method presented in Ref. \cite{KGHZ} gives necessary and
sufficient conditions for the violation of local realism which
strongly suggests that the inequality found has similar property.

\end{document}